      [1999/12/01 v1.4c Il Nuovo Cimento]
\documentclass{cimento}

\newif\ifppn
\ppntrue

\usepackage{graphicx}  
\usepackage{pstricks}
\title{
NLO vector boson production with light jets}
\author{
Z.~Bern\from{ins:UCLA}\ETC,
G.~Diana\from{ins:Saclay},
L.~J.~Dixon\from{ins:SLAC},
F.~Febres Cordero\from{ins:Caracas},
D.~Forde\from{ins:cern}\from{ins:nikhef},
T.~Gleisberg\from{ins:SLAC},
S. H\"oche\from{ins:SLAC},
H. Ita\from{ins:UCLA}\from{ins:NB},
D.~A.~Kosower\from{ins:Saclay},
D.~Ma\^{\i}tre\from{ins:cern}\from{ins:ippp} \atque
K.~Ozeren\from{ins:UCLA}
}
\instlist{
\inst{ins:UCLA} Department of Physics and Astronomy, UCLA, Los Angeles, CA 90095-1547, USA
\inst{ins:Saclay} Institut de Physique Th\'eorique, CEA--Saclay, F--91191 Gif-sur-Yvette cedex, France
\inst{ins:SLAC} SLAC National Accelerator Laboratory, Stanford University, Stanford, CA 94309, USA 
\inst{ins:Caracas} Departamento de F\'{\i}sica, Universidad Sim\'on Bol\'{\i}var,  Caracas 1080A, Venezuela
\inst{ins:cern} PH-TH, Case C01600, CERN, CH-1211 Geneva 23, SWITZERLAND 
\inst{ins:nikhef}NIKHEF Theory Group, Science Park 105, NL--1098~XG  Amsterdam, The Netherlands
\inst{ins:NB} Niels Bohr International Academy and Discovery Center, NBI, DK-2100 Copenhagen, DK
\inst{ins:ippp} Institute for Particle Physics Phenomenology, Ogden Centre for Fundamental Physics, Department of Physics, University of Durham, Science Laboratories, South Rd, DURHAM DH1 3LE, United Kingdom
}
\PACSes{
\PACSit{12.38.Bx}{Perturbative calculations}
\PACSit{14.70.Fm}{W bosons}
\PACSit{14.70.Hp}{Z bosons}
}

\begin{document}

\maketitle

\begin{abstract}
\ifppn
\rule{1cm}{0cm}\vspace{-10cm}\\\hbox{\rm\small
UCLA/12/TEP/101$\null\hskip 1.5cm \null$
SLAC--PUB--14855$\null\hskip 1.5cm \null$
CERN--PH--TH/2012/013
}
\hbox{\rm\small $\hskip 4.7cm \null$ IPPP/12/02 }
\rule{1cm}{0cm}\vspace{9.5cm}\\
\fi
In this contribution we present recent progress in the computation of
next-to-leading order (NLO) QCD corrections for the production of an
electroweak vector boson in association with jets at hadron colliders. 
We focus on results obtained using the virtual matrix element
library \texttt{BlackHat}~\cite{BlackHatI} in conjunction with
\texttt{SHERPA}~\cite{Sherpa,Amegic,Amegic2}, focusing on results relevant to
understanding the background to top production.
\end{abstract}

\section{Introduction}
The production of a vector boson in association with several jets at
the Large Hadron Collider (LHC) is an important background for other
Standard Model processes as well as new physics signals. In
particular, the production of a $W$ boson in association with many
jets is an important background for processes involving one or more
top quarks.

Precise predictions for the backgrounds are crucial to measurement of
top-quark processes.  Vector boson production in association with
multiple jets is also a very important background for many SUSY
searches, as it mimics the signatures of many typical decay chains.
Here we will discuss how polarisation information can be used as an
additional handle to differentiate top pair production from ``prompt''
$W$-boson production~\cite{wpol}.  More generally, ratios of
observables, for example for events containing a $W$ boson versus
those containing a $Z$ boson, are expected to be better-behaved as
many uncertainties cancel in such ratios.
Precise calculation of ratios, along with measurement of one of the
two processes in the ratio, can be used in data-driven techniques for
estimating backgrounds.

\section{Recent results}


In this contribution we present NLO results obtained by combining
results from two programs, {\tt SHERPA} \cite{Sherpa,Amegic,Amegic2} for the
real emission and {\tt BlackHat} \cite{BlackHatI} for the virtual
emission. {\tt SHERPA} is also used to perform the integration over
the phase space of the virtual contribution.  We also used {\tt
BlackHat} to supply the most complex tree-level matrix elements needed
for the real emission contributions.

The calculation of virtual corrections to complex processes four of
more final state objects was for a long time a highly non-trivial
challenge.  This situation has changed with the development of
unitarity-based techniques (for a review see \cite{OnShellReview,OnShellReview2}).
In particular, the inclusive production of a $W$ boson in association
with three jets has been computed at NLO by two independent groups
with these techniques using different color
approximations \cite{PRLW3BH,EMZW3Tev} followed by a full color
computation~\cite{W3jDistributions}. The same process with the $W$
boson replaced by a $Z$ boson, at the Tevatron, was computed shortly
thereafter~\cite{TeVZ}.  More recently, NLO results
for the production of a $W$ or a $Z$ boson in association with four
jets have been presented \cite{BHSW4,BHSZ4}.  These are the first NLO
QCD calculations at hadron colliders involving five final state
objects, including jets.

Fig.~1~\cite{BHSZ4} shows the transverse momentum
distributions of the first, second, third and fourth jets in events
with a $Z$ boson and at least four jets. In the top row of plots, the
dashed (blue) curve represents the LO result while the plain (black)
line is the NLO result. The middle part of the plot shows the ratio to
the NLO prediction. The bands display the scale variation obtained by
varying the renormalisation and factorisation scale simultaneously by
factors of $1/2$, $1/\sqrt{2}$, $1$, $\sqrt{2}$ and $2$.  Clearly, the
scale variation for the NLO result (in gray) is much smaller than that
attributed to the LO result (in hashed orange).  The bottom panels in 
this figure show the ratios between the $W$ and $Z$ boson
processes. These ratios and their stability going from LO to NLO are
important as they are used by experimenters to
extrapolate from a measured control process into a signal process or
region.
\begin{figure}\label{fig:Z4}
\includegraphics[scale=0.43,clip]{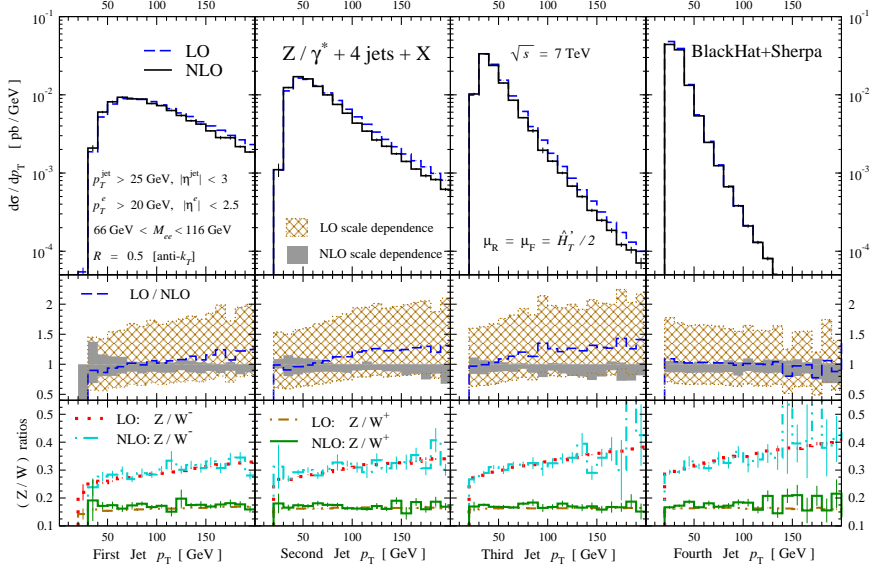}
\caption{$P_T$ distributions for the first four jets in $Z$+4 jets at the LHC with $\sqrt{s}=7$~TeV, as well as ratios to $W^+$ and $W^-$.  
Details of the experimental cuts can be found in ref.~\cite{BHSZ4}.}
\end{figure}

\section{$W$ polarisation at the LHC}
The fact that $W$ bosons produced at low transverse momentum, moving
mainly along the beamline, are predominantly polarised left-handed has
been known for a long time~\cite{ESW}.  More recently it has been
recognized that the same is true for $W$ bosons produced at large
transverse momentum, and the mechanism has been described as
well~\cite{wpol}.  In contrast to the longitudinal effect which relies
on angular momentum conservation, the transverse effect is more subtle
and may be understood as properties of the matrix elements.  The
effective polarisation has been found to be relatively unaffected by
NLO corrections~\cite{wpol}.  This polarisation effect can be used as
an additional handle to separate $W$ bosons directly produced in
conjunction with jets from those produced in the decay of a top or
anti-top quark, as the latter are predominantly longitudinally
polarised.

A left-handed $W^+$ that decays leptonically will preferentially emit the neutrino along its flight direction, whereas the positron will tend to be emitted in the opposite direction (in the $W$ rest frame), resulting in a larger average transverse momentum for the neutrino than for the positron. The opposite effect happens in the case of the decaying left-handed $W^-$ boson: the electron inherits a larger average transverse momentum than the anti-neutrino. This asymmetry in the transverse momentum of the decay products is not present if the vector boson is longitudinally polarised. Fig.~2 illustrates the phenomenon in density plots for the ratio of $W$s arising from top-quark decays to those from prompt production  at the LHC $\sqrt{s}=7$ TeV and cuts chosen as in ref. \cite{wpol}:
\[\frac{{\rm d}^2\sigma}{{\rm d}p_T^\nu{\rm d}p_T^l}\left(pp\rightarrow t\bar t\rightarrow W^\pm+\mbox{3 jets}\right)\bigg/\frac{{\rm d}^2\sigma}{{\rm d}p_T^\nu{\rm d}p_T^l}\left(pp\rightarrow W^\pm+\mbox{3 jets}\right)\;.\]
One can see in the figure that in the case of the $W^+$ bosons, prompt production tends to populate more the region where the transverse momentum of the neutrino is larger than that of the positron, while the opposite effect is seen for the $W^-$ boson. A suitable cut, or the addition of such information in statistical analysis tools such as boosted decision trees or neural networks could help discriminate between these two processes.       
\begin{figure}\label{fig:Wtt}
\begin{pspicture}(0,-2.8)(9.0,4.0)
\rput{90}(3.0,0.7){\includegraphics[scale=0.35]{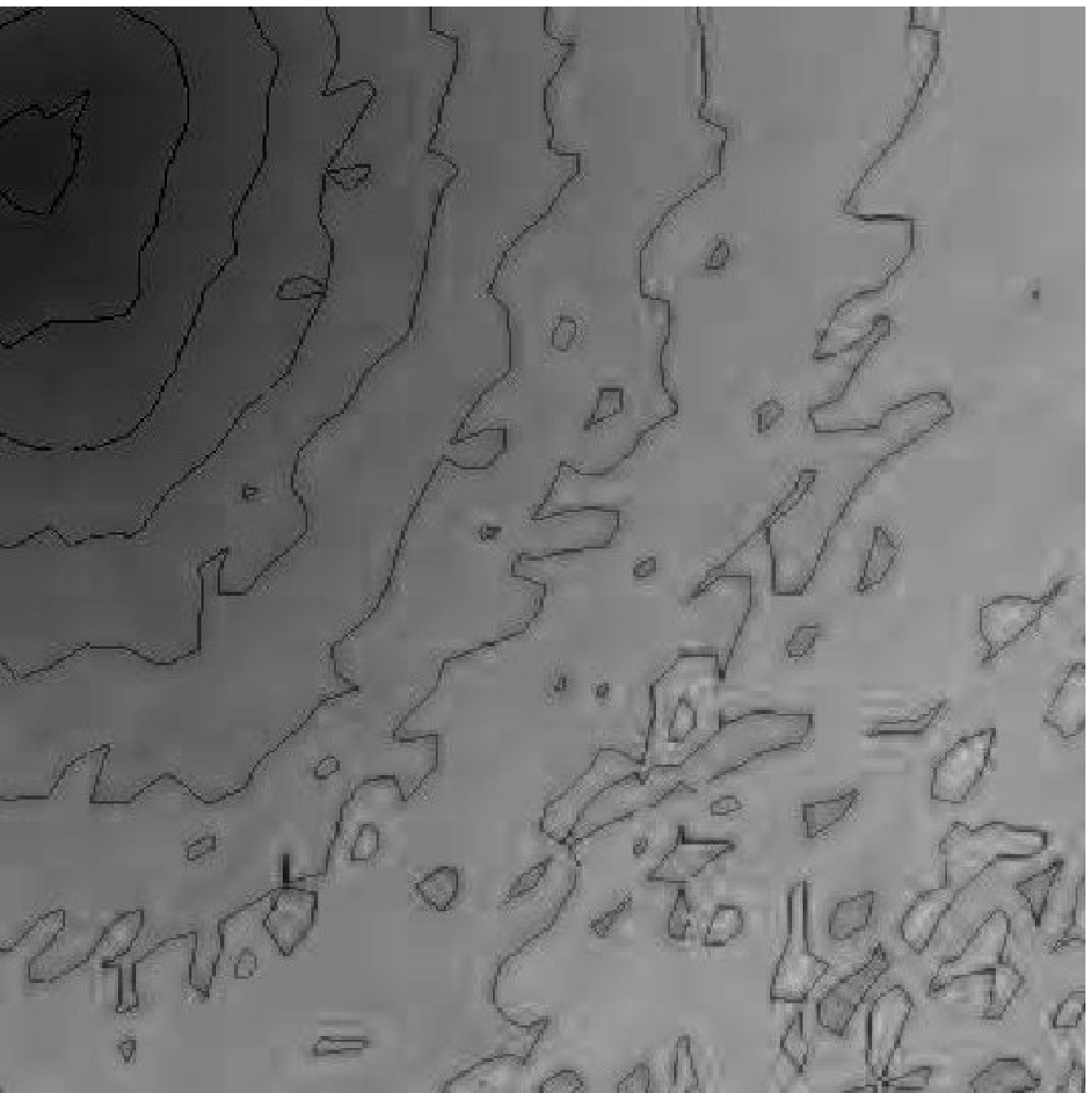}}
\usefont{T1}{ptm}{m}{n}
\rput(3.3,3.6){\Huge $W^+$}
\psline[linewidth=0.04cm,arrowsize=0.1cm 2.0,arrowlength=2.4,arrowinset=0.0]{->}(0.30,-2.0)(0.3,3.7)
\psline[linewidth=0.04cm,arrowsize=0.1cm 2.0,arrowlength=2.4,arrowinset=0.0]{->}(0.30,-2.)(5.80,-2.0)
\usefont{T1}{ptm}{m}{n}
\rput(5.6,-2.4){\large $p_T^{e^+}$}
\usefont{T1}{ptm}{m}{n}
\rput(-0.15,3.25){\large $p_T^\nu$}
%
%
%
\rput(9.0,3.6){\Huge $W^-$}
\rput{90}(9.0,0.7){\includegraphics[scale=0.35]{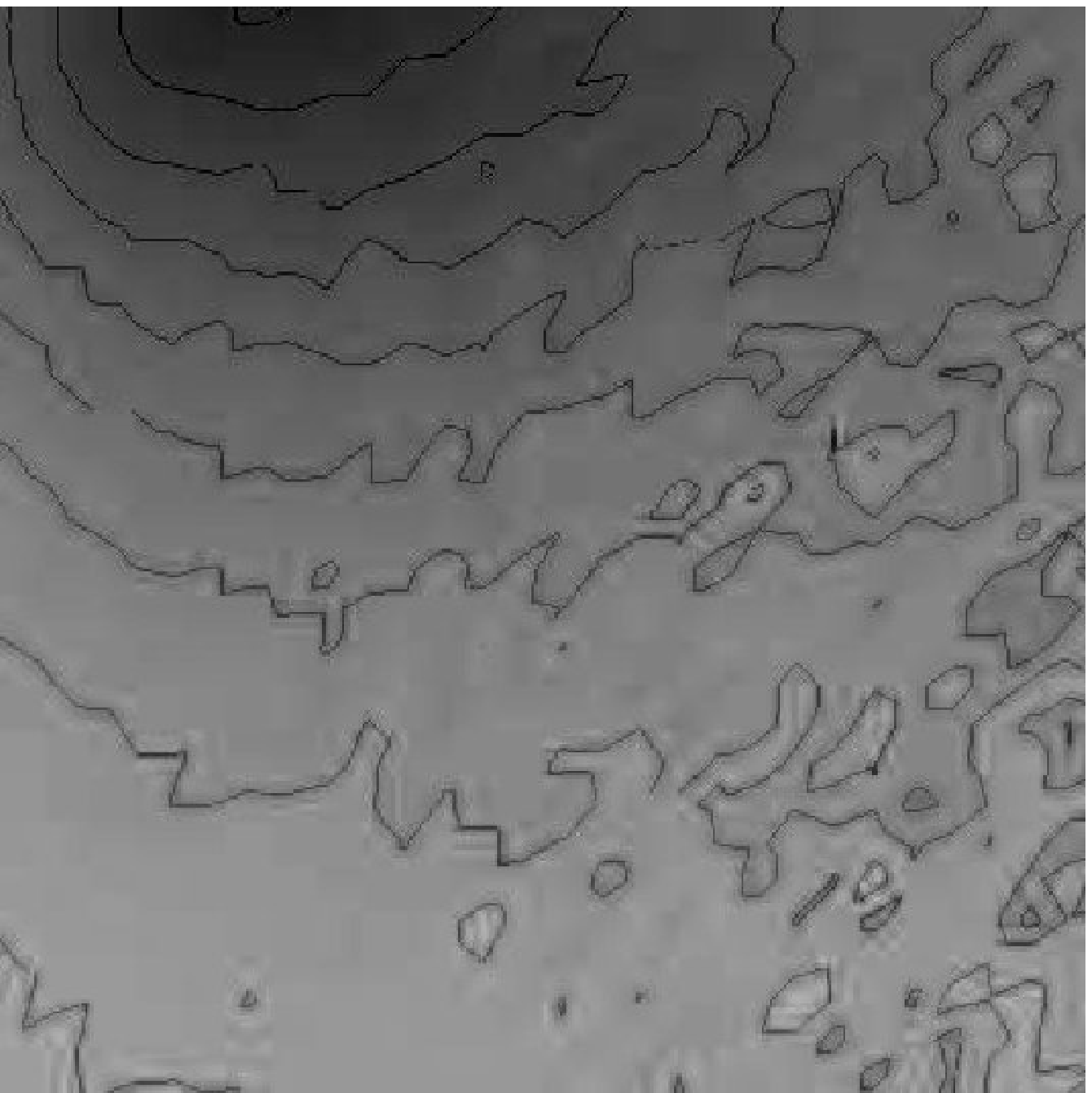}}
\usefont{T1}{ptm}{m}{n}

\psline[linewidth=0.04cm,arrowsize=0.1cm 2.0,arrowlength=2.4,arrowinset=0.0]{->}(6.30,-2.0)(6.3,3.7)
\psline[linewidth=0.04cm,arrowsize=0.1cm 2.0,arrowlength=2.4,arrowinset=0.0]{->}(6.30,-2.)(11.80,-2.0)
\usefont{T1}{ptm}{m}{n}
\rput(10.6,-2.4){\large $p_T^{e^-}$}
\usefont{T1}{ptm}{m}{n}
\rput(5.85,3.25){\large $p_T^{\bar{\nu}}$}

\rput{90}(12.5,1.0){\includegraphics[scale=0.15]{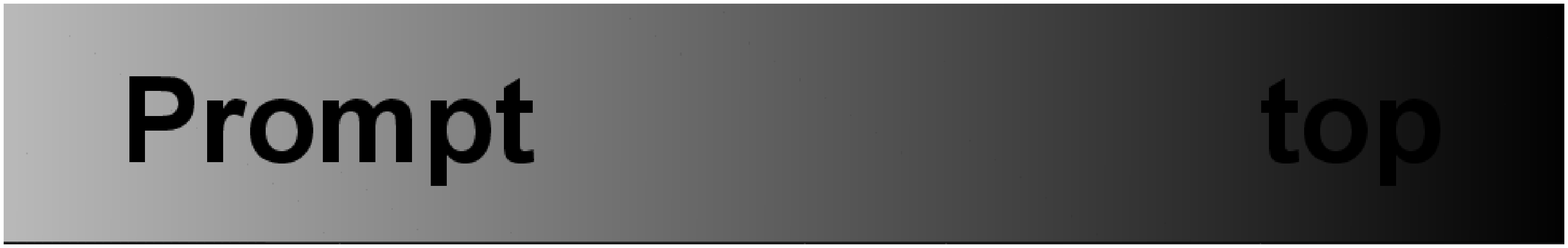}}

\end{pspicture} 
\caption{Density plots for the ratio of the LO cross sections for $pp\rightarrow t\bar t\rightarrow W^\pm+\mbox{3 jets} $ and $pp\rightarrow W^\pm+\mbox{3 jets}$ as a function of the lepton and neutrino transverse momentum.}
\end{figure}
The predominant left-handedness of prompt $W$ bosons at the LHC with
$p_T>50$~GeV has been measured by the CMS
collaboration \cite{CMSWpolpaper} and found to be in good agreement
with theoretical predictions~\cite{wpol}.

\section{Conclusions}
The field of NLO corrections for high-multiplicity final states has
seen tremendous progress in the last few years, as exemplified by the
first computations of $2\rightarrow 5$ processes. This development is
very timely as precise predictions are helpful for LHC
measurements. In particular, as we discussed here, the left-handed
polarisation of prompt $W$ bosons in association with jets
at the LHC may be useful as an additional handle to separate them from
$W$ bosons produced in the decay of heavy particles such as top quarks.
\acknowledgments
This work has been supported by the European Research Council under Advanced Investigator Grant ERC--AdG--228301, and by the US Department of Energy under contract DE–AC02–76SF00515. SH is supported in part by the US National Science Foundation, grant NSF–PHY–0705682, (The LHC Theory Initiative). DM’s work was supported by the Research Executive Agency (REA) of the European Union under the Grant Agreement number PITN-GA-2010-264564 (LHCPhenoNet). H.I.’s work is supported by a grant from the US LHC Theory Initiative through NSF contract PHY–0705682.

\end{document}